%% file: main.tex
\RequirePackage[T1]{fontenc}
\documentclass[conference]{IEEEtran}
\usepackage{amsmath,amssymb,booktabs,array,graphicx}
\usepackage[hidelinks]{hyperref}
\input{numbers}
\providecommand{\PHnote}[1]{}
\providecommand{\EHidGptACov}{58}\providecommand{\EHidGptAReg}{12}\providecommand{\EHidGptRCov}{66}\providecommand{\EHidGptRReg}{7}
\providecommand{\EHidGptBCov}{64}\providecommand{\EHidGptBReg}{8}\providecommand{\EHidGptCCov}{74}\providecommand{\EHidGptCReg}{3}
\providecommand{\EHidCACov}{61}\providecommand{\EHidCAReg}{11}\providecommand{\EHidCRCov}{69}\providecommand{\EHidCRReg}{6}
\providecommand{\EHidCBCov}{67}\providecommand{\EHidCBReg}{7}\providecommand{\EHidCCCov}{79}\providecommand{\EHidCCReg}{2}
\providecommand{\EHidCPGptCov}{52}\providecommand{\EHidCPGptReg}{0}
\providecommand{\EHidGptPGptCov}{49}\providecommand{\EHidGptPGptReg}{0}
\providecommand{\EFourContexts}{648}\providecommand{\EFourPairs}{1{,}024}
\providecommand{\EFourFamilyFlips}{18}\providecommand{\EFourInjected}{419}

\providecommand{\EFourFamilyGain}{21}\providecommand{\EFourFamilyCI}{11 to 30}
\providecommand{\EFiveTrainGPUh}{46}\providecommand{\EFiveLocalPerSearchUSD}{0.3}\providecommand{\EFiveFrontierPerSearchUSD}{1.1}\providecommand{\EFiveBreakEven}{144}
\ifdefined\XeTeXversion\renewcommand{\texttt}[1]{\textit{#1}}\fi
\makeatletter
\def\@IEEENORMtitlevspace{0.25\baselineskip}% author block to body
\def\@IEEEMINtitlevspace{0\baselineskip}
\makeatother
\AtBeginDocument{\setlength{\abovedisplayskip}{3pt}\setlength{\belowdisplayskip}{3pt}\setlength{\abovedisplayshortskip}{1pt}\setlength{\belowdisplayshortskip}{3pt}}% compact display math
\newcommand{\Sone}{\mathrm{S1}}
\newcommand{\Stwo}{\mathrm{S2}}

\title{The Code Behind the Layout: AI Agents for Analog Layout Generators}
\author{\IEEEauthorblockN{Minxing Chu\textsuperscript{1}, Anhang Li\textsuperscript{2}, Martins Atilla\textsuperscript{1}, Ruichen Qi\textsuperscript{1}, Mehdi Saligane\textsuperscript{1}}
\IEEEauthorblockA{\textsuperscript{1}Brown University, Providence, RI, USA\quad \textsuperscript{2}University of Michigan, Ann Arbor, MI, USA\\
\{minxing\_chu, martins\_atilla, ruichen\_qi, mehdi\_saligane\}@brown.edu, anhangli@umich.edu}}
\begin{document}
\maketitle
\suppressfloats[t]% keep the first column for the abstract; Fig. 1 then opens the right column

\begin{abstract}
\looseness=-1
Analog layout remains largely manual and hard to automate. Generator frameworks such as gLayout express layouts as programs, producing layouts from parameterized, design-rule-aware generators; however, writing such generators still takes expert effort. Large language models (LLMs) could now write generator code, and frontier models can even draw layout geometry from design intent. This raises a practical question: if a model can draw the layout itself, what does a generator framework still contribute, and is applying these models to a generator framework better than having them draw a layout from scratch? Through controlled experiments on gLayout in SKY130, we find that the main obstacle to drawing a layout directly is its time and token cost, and that the main advantage of a model building on a generator framework is reusability. Its verified primitives keep each model edit small enough to check, and its parameters allow a single edit to be evaluated across an entire family of layouts. Accordingly, a model is most effective when it develops the layout from the generator source of the family that the design intent belongs to, and tests each edit on the family members most likely to fail, so that one verified change improves every layout the generator produces. The results of these family tests can then serve as supervision for a local model, which learns edits that hold across the family at a fraction of the inference cost.
\end{abstract}

\section{Introduction}
\begin{figure}[t]
  \centering
  \includegraphics[width=0.98\columnwidth]{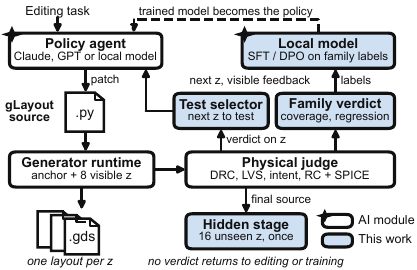}
  \vspace{-6pt}
  \caption{Overview of the family-level editing loop.}
  \label{fig:workflow}
  \vspace{-5pt}
  \vspace{1pt}
\end{figure}

\looseness=-1
Analog layout remains one of the least automated stages of integrated circuit design.
A current mirror laid out in one process is typically redrawn for every new sizing and every new technology node.
The reasoning behind the layout, such as which devices must match, which nets carry signal, and where taps and routes are placed, is rarely recorded and must be reconstructed each time.
Generator-based frameworks capture this reasoning in code.
The BAG family expresses schematics, layouts and measurements as Python generators~\cite{guo2025bag3}, and LAYGO2 assembles layouts from dynamic templates on technology grids~\cite{shin2023laygo2}.
ALIGN and MAGICAL derive placement and routing from netlists and constraints~\cite{kunal2019align,xu2019magical}, while gLayout composes hierarchical device and routing generators over a mapped PDK~\cite{hammoud2024language,hammoud2024generator} (Table~\ref{tab:interfaces}).
In each case, a generator is a parameterized program: a new set of device sizes yields a new member of the same layout family, and the designer's decisions carry over to every member.

\looseness=-1
However, writing a generator still requires substantial expertise in both the layout itself, from design-rule spacing and device construction to net routing, and the framework's programming interface used to express it.
Meanwhile, large language models (LLMs) have become increasingly capable of solving non-trivial programming tasks~\cite{jimenez2024swebench,yang2024sweagent}, and researchers have begun to apply them to generator design and to layout itself~\cite{saligane2026generative}.
LLMs now read circuit schematics~\cite{akbari2026circuitsense}, propose circuit topologies~\cite{chang2024lamagic,gao2025analoggenie} and size analog circuits~\cite{yu2026autosizer}.
They also translate design intent into generator code~\cite{hammoud2024language}, extend existing generators, tune their exposed parameters from post-layout feedback~\cite{liu2025layoutcopilot,liu2026simulation}, and repair design-rule violations with a checker in the loop~\cite{mukherjee2026drcaid,chu2026trace}.
Moreover, state-of-the-art models can even draw an entire layout from scratch, polygon by polygon, directly from design intent, without a generator backbone in the loop~\cite{liu2025layoutcopilot}.

\looseness=-1
This progress raises a practical question with two sides.
If a model can produce a layout by itself, why keep a generator framework at all?
And if the framework stays, how much does the model gain by working inside it, editing and extending the generator?
We study this question from three aspects.
The first is at which level the model should act.
The three levels in Table~\ref{tab:interfaces}, the raw geometry, the calls to the framework's device and routing cells, and the generator source, differ in where the design knowledge resides and in how much code an edit has to touch, and a policy that cannot land a working edit at a level gains nothing from it.
The second is how an edit should be evaluated.
Checking it only on the set of device sizes it was written for, which we call a configuration, is what existing layout agents do; checking it across the family is what the generator makes possible, and the question is how many configurations that check needs.
The third is whether the verdicts of those checks can train a smaller local model to do the same work at lower cost.
Cost has two sources, the physical verification each tested configuration requires and the inference tokens the frontier model consumes; a verdict that holds across a family can serve as a training label and so reduce the second.

\looseness=-1
We address these questions through controlled experiments on gLayout~\cite{hammoud2024generator,openfasoc} in SKY130.
Two frontier models, Claude Opus~5.5 and GPT-6.1, serve as the editing policy: the agent that proposes the edits.
Each runs the same tasks at three action levels: raw geometry, device and routing cells, and full generators.
A local 32B model trained on the family verdicts then takes over the editing (Fig.~\ref{fig:workflow}).

\begin{table*}[t]
  \centering
  \caption{Where layout frameworks keep their knowledge and what a policy can change.}
  \label{tab:interfaces}
  \footnotesize
  \vspace{-5pt}
  \renewcommand{\arraystretch}{0.95}
  \begin{tabular}{p{.19\textwidth}p{.375\textwidth}p{.355\textwidth}}
    \toprule
    Interface & Location of design knowledge & Policy action \\
    \midrule
    Raw GDS geometry~\cite{gdstk} & Program-built polygons, layers, labels and nets & Edit geometry construction and net routing \\
    Primitive Python API~\cite{hammoud2024generator} & Device and via construction with port interfaces & Place primitive instances and connect terminals \\
    gLayout~\cite{hammoud2024language,hammoud2024generator} & MappedPDK, hierarchical device and block generators & Edit generator source or compose allowlisted blocks \\
    BAG/BAG2/BAG3++~\cite{guo2025bag3} & Procedural schematic and layout generators, test scripts & Edit generator procedures and design parameters \\
    LAYGO2~\cite{shin2023laygo2} & Dynamic templates, technology grids, reusable instances & Edit template generation, grid placement and routing \\
    ALIGN~\cite{kunal2019align} & Netlist hierarchy, PDK abstraction, inferred constraints & Supply circuit- and hierarchy-level constraints \\
    MAGICAL~\cite{xu2019magical} & Constraint generation, placement and routing engines & Supply circuit intent and symmetry constraints \\
    \bottomrule
  \end{tabular}
  \vspace{-6pt}
\end{table*}

\section{Tasks, Verification and Setup}\vspace{-3pt}
\label{sec:contract}
\looseness=-1
All three aspects are evaluated on the same tasks and with the same notion of a passing layout stated in this section.
We denote a generator by $G_\theta(z,p)$, where $\theta$ is its editable source code, $z$ a legal configuration of the circuit (its device sizes and finger counts), and $p$ the fixed process.
The family of $\theta$ is the set of configurations it serves.
A patch is a source edit proposed by a policy; applying it to the original source $\theta_0$ yields a revised source $\theta$.
For every $z$, the task fixes a read-only reference: the netlist $N(z,p)$; the intended devices with their widths, lengths and finger counts; and the external ports.
The task also specifies how the layout is evaluated: a testbench with supply, stimulus and load; the specifications, each with a direction and a limit; and the process corners.
Because the sizing is contained in $z$, a patch can only change how a layout is constructed, without modifying device size, ports, or the reference.

\looseness=-1
A candidate can fail at several stages, from generator code that does not execute to a clean layout that misses one specification at one corner, and each stage further is more costly to reach.
We define two tiers that record the last stage a candidate passed, so that outcomes remain comparable across tasks.
$\Sone$ requires that the generated layout pass full-rule DRC, LVS against $N(z,p)$, and a device check that compares the layout's devices and terminals with the intended ones finger by finger, including each finger's model, $W/L$ and multiplicity and each port's identity and role, so that a matching total width alone cannot pass.
$\Stwo$ additionally requires parasitic extraction, with all extracted elements physically present in the layout, and converged simulations that meet every specification at the typical (tt), slow (ss) and fast (ff) process corners at 27$^\circ$C and 1.8\,V.
We call a candidate feasible on a configuration if it reaches $\Stwo$ there, and we call the extraction and three-corner simulation that $\Stwo$ requires a full evaluation.

\looseness=-1
Quality is compared only among feasible candidates, using one objective per circuit class: layout area for the component generators, and worst-corner unity-gain bandwidth (UGB) per unit area for the operational transconductance amplifiers (OTAs).
Feasibility is a constraint rather than a weighted term, so an unstable amplifier cannot be rewarded for its bandwidth.

\looseness=-1
All experiments run on 24 tasks covering eight generators, six component generators (a current mirror, a cascode mirror, a source follower, a stacked switch, a common-source stage and a resistively loaded pair) and two OTAs (the five-transistor and compact Miller amplifiers); each task starts from a layout that misses an electrical target, breaks a connection or leaves room to improve.
Each task declares its family: the configuration it was written for, which we call the anchor, eight further visible configurations the policy may test, and sixteen hidden configurations, inside and beyond the visible range, evaluated once at the end with no feedback to the policy.
Two frontier policies, Claude Opus~5.5 and GPT-6.1, run every task.
Each episode, which we also call a search, is a fresh session: the policy submits each patch as a git-style unified diff and requests evaluations; each reply is limited to 1,024 tokens.
Every search records its output tokens and its physical CPU time, so each result comes with its cost.
Sections~\ref{sec:level} to \ref{sec:local} take the three aspects of the introduction in turn.

\section{Where to Act: Generator Edits Cost Least}\vspace{-3pt}
\label{sec:level}
\begin{figure}[t]
  \centering
  \includegraphics[width=\columnwidth]{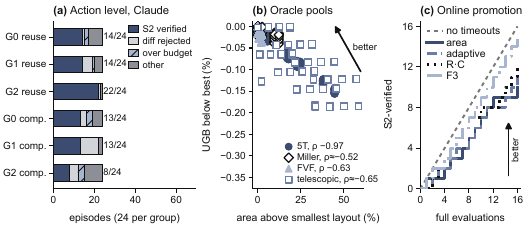}
  \vspace{-16pt}
  \caption{Action level and evaluation fidelity shape the search. (a) Outcome of every Claude Opus~5.5 episode by action level (G0 geometry, G1 primitives, G2 generator source) and by whether the needed cell could be reused or had to be built from primitives. (b) Oracle pools of the four amplifiers (5T: five-transistor, FVF: flipped voltage follower) with the Spearman correlation between area and bandwidth. (c) Candidates verified within the time limit against full evaluations on a 5T pool whose layouts all pass, under four screens: area, adaptive geometric rule, extracted RC product and calibrated interval F3.}
  \label{fig:abstraction}
\end{figure}
\looseness=-1
The action-level study runs every task at three levels: the policy writes raw geometry (G0), places primitive devices and wires (G1), or edits the generator source (G2).
In half of the tasks the needed cell exists as a complete library cell that the policy may reuse; in the other half that cell is withheld and the policy must build it from the framework's lower-level primitives, individual devices, vias and routes.

\looseness=-1
For GPT-6.1, whose prompt carried no rules about the format of its replies, only one of the three levels works.
It verified one episode out of \EOneGZeroN{} when it drew geometry (G0) and one out of \EOneGOneN{} when it placed primitives (G1); nearly every other episode ended because the first diff did not apply or the reply ran past the 1,024-token limit.
At the generator level (G2) the same policy verified \EOneGTwoReuseS{} of \EOneGTwoReuseN{} reuse episodes within a median of \EOneGTwoMedianDecisions{} decisions, and \EOneSuccessHiddenFull{} of all \EOneSuccessTotal{} anchors it verified held on every hidden configuration.
The size of the edit explains the gap.
A first diff at G0 or G1 added \EOneGZeroFirstDiffLines{} to \EOneGOneFirstDiffLines{} lines and a decision cost about 2,000 output tokens including reasoning; at G2 a decision cost \EOneGTwoOutTok{}, because changing one argument or one call is usually enough.
Claude Opus~5.5, whose prompt carries format rules, lands edits at every level (Fig.~\ref{fig:abstraction}(a)) and shows the same pattern: \ECOneHiddenFull{} of its \ECOneSuccessTotal{} verified anchors held on every hidden configuration.
The generator's verified building blocks are what keep the edit small: without the needed cell, GPT-6.1 almost always fails and Claude Opus~5.5 does worst at G2.

\looseness=-1
Frontier inference, counted per verified configuration, favors the generator level.
At the geometry and primitive levels almost nothing GPT-6.1 wrote passed, while at the generator level most reuse episodes verified, and each verified edit was evaluated on its sixteen hidden configurations as well, so one edit could yield seventeen verified configurations and a verified configuration cost \EOneLowToGTwoTokRatio{} times fewer output tokens there: about \EOneGTwoTokPerVerifiedK{}k output tokens per verified configuration at the generator level against \EOneLowTokPerVerifiedK{}k below it.
For Claude Opus~5.5, which lands edits at every level, the ratio is still \ECOneLowToGTwoTokRatio{} to one.
The physical CPU time per verified configuration is similar at every level, about \EOneGTwoCpuPerVerifiedMin{} minutes at the generator level and \EOneLowCpuPerVerifiedMin{} below it, so the saving lies in inference: the tokens the model generates and the time they take.

\section{Evaluating an Edit: Judge It on the Family}\vspace{-3pt}
\label{sec:family}
\begin{figure*}[t]
  \centering
  \includegraphics[width=\textwidth]{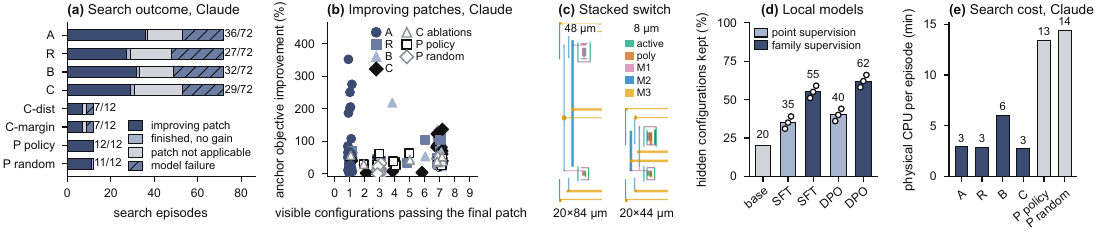}
  \vspace{-18pt}
  \caption{Editing searches, their evidence, the local models and their cost. (a) Outcome of every Claude Opus~5.5 search by test rule (A anchor only, R random, B boundary, C risk; C-dist and C-margin keep one term of C) and parameter baseline (P policy, P random). (b) Anchor improvement against visible configurations passing the final patch. (c) The stacked switch before and after a one-line change. (d) Hidden configurations kept by the local models under point and family supervision, SFT and DPO, three seeds. (e) Median physical CPU per search episode.}
  \label{fig:editing}
  \vspace{2pt}
\end{figure*}
\looseness=-1
An edit at the generator level changes the source that every configuration shares, so it reaches the whole family at once; that is where its value comes from, and also its risk.
In the level study above, most verified edits held on all sixteen hidden configurations, but some passed the configuration they were written for and broke another, and a check on one configuration cannot tell the two apart.
We therefore score an edit on a set $D$ of configurations by two quantities.
Coverage is the fraction of configurations in $D$ that reach $\Stwo$ under the revised source $\theta$.
Regression is the fraction of configurations that reached $\Stwo$ under the original source $\theta_0$ but no longer do under $\theta$:
\begin{align}
  \mathrm{Coverage}_D(\theta) &= |D|^{-1}\textstyle\sum_{z\in D}\mathbf 1[\Stwo(\theta,z,p)],\\
  \mathrm{Regression}_D(\theta,\theta_0) &= \frac{\sum_{z\in D}\mathbf 1[\Stwo(\theta_0,z,p)\land\neg\Stwo(\theta,z,p)]}{\sum_{z\in D}\mathbf 1[\Stwo(\theta_0,z,p)]}.
\end{align}
Two questions follow: how much physical work the check of one configuration needs, and which configurations to check when the budget covers only part of the family.

\vspace{-2pt}
\subsection{Checks need the physical stages, rarely the full simulation}
\vspace{-2pt}
\looseness=-1
The fidelity study asks whether a cheaper signal can replace the extraction and three-corner simulation of a full evaluation.
We use four OTAs, the five-transistor, compact Miller, flipped voltage follower and telescopic amplifiers, each at three sizings.
For each sizing the generator can place and route the circuit in 64 legal ways, giving 64 layouts.
We run each of them through the full evaluation, and use results as the oracle.
We then test two cheap signals against this oracle: the bounding-box area of the layout, and
$\mathrm{RC}_\Sigma$, the product of the total extracted resistance and the total extracted capacitance~\cite{chu2026trace}.
We also test four screening strategies~\cite{kandasamy2017multifidelity}.
Each strategy ranks the 64 layouts and sends the top 16 to full evaluation, ranking by area, by an adaptive geometric rule, by $\mathrm{RC}_\Sigma$, or by a calibrated interval F3, which predicts a bound on worst-corner performance from a single nominal-corner simulation.

\looseness=-1
Verification has two separate jobs, deciding whether a layout passes and deciding which passing layout is best, and the two need different amounts of physics.
Every layout in the oracle pools of Fig.~\ref{fig:abstraction}(b) went through the whole flow, and every layout that passed DRC, LVS and the device check also met every specification at all three corners.
Deciding which passing layout is best hardly matters: within a pool, the layout choices move the worst-corner bandwidth by less than a fifth of a percent.
On the five-transistor OTA, the bounding-box area ranks the passing layouts almost perfectly (Spearman $\rho=\ETwoSZeroSpearman$); on the compact Miller OTA it ranks loosely, but the bandwidth it misses amounts to \ETwoMillerSZeroUgbSpread\%.
The cheap signals fail at the first job, predicting the pass itself: three quarters of the Miller layouts fail the physical checks although every one of them is legal, ranking by geometry sends exactly those on to full evaluation, and only \ETwoOnMillerSZeroAreaSTwo{} of the 16 Miller layouts selected by area passed.
Every Miller layout selected by the post-extraction signals $\mathrm{RC}_\Sigma$ and F3 passed.
On a five-transistor pool every layout passes given enough simulation time, so a screen loses only evaluations that exceed the time limit, and $\mathrm{RC}_\Sigma$ and F3 lost the fewest (Fig.~\ref{fig:abstraction}(c)).
Thus, we conclude that a check should run the physical stages on every candidate but the three-corner simulation only on the candidates that survive extraction; this is how every full evaluation in the rest of the paper is run.

\vspace{-2pt}
\subsection{Test the configurations where an edit is most likely to fail}
\vspace{-2pt}
\looseness=-1
The editing study asks which configurations to test a patch on when the eight full evaluations of a search cannot cover the family.
Four rules spend the same budget~\cite{sketch2006,mechtaev2016angelix}: A tests the anchor only, R draws visible configurations at random, B tests the edges of the visible range, and C tests the untested configuration with the largest sum of its distance to those tested on the current patch and a score that rises as its specification margin falls, the margin being the worst one measured so far at the nearest configuration; C-margin and C-dist keep only one term.
A parameter-only baseline, P, tunes the layout options the generator already exposes, such as spacings and routing choices, at random or with simulation feedback~\cite{liu2026simulation}.

\looseness=-1
Each search ends with one revised generator, and we check whether that revision holds on the other configurations of the same circuit (Table~\ref{tab:e3}).
Most revisions do.
The eight full evaluations of a search cover the anchor and seven of the eight other visible configurations, and three quarters of the revisions that improved the anchor's objective under rules R, B and C passed all seven (Fig.~\ref{fig:editing}(b)).
The stacked switch in Fig.~\ref{fig:editing}(c) shows why a shared edit pays off: a one-line change to the generator source halves the area of the anchor layout, and because every configuration is built by the same source, the layouts of the other configurations of the switch shrink as well.
Across tasks, source edits reached up to \ECThreeCompImpMax\% less area for the component generators and \ECThreeOtaImpMax\% more bandwidth per area for the OTAs.
The regressions that did occur are exactly the cases an anchor-only test cannot see: on the compact Miller OTA, \EThreeSiblingFailures{} GPT and \ECThreeSiblingFailures{} Claude revisions passed every check on the anchor while another configuration failed DRC, whereas on the component generators every revision that passed its anchor held everywhere.
On the sixteen hidden configurations, evaluated once at the end, rule C keeps the most configurations working with the least breakage and rule A the fewest, for both policies.
Screening every candidate with the calibrated interval F3 before testing it on the family, the variant C+F3, also finds strong amplifier revisions, but at about twenty times the physical cost of a rule-C search.

\begin{table}[t]
  \centering
  \caption{Editing searches by test rule and policy: searches (\textit{n}), improving patches, their median gain (\%), and hidden coverage and regression (\%).}
  \label{tab:e3}
  \footnotesize
  \setlength{\tabcolsep}{2.2pt}
  \vspace{-5pt}
  \renewcommand{\arraystretch}{0.97}
  \begin{tabular}{lrrrrrrrrrr}
    \toprule
    & \multicolumn{5}{c}{GPT} & \multicolumn{5}{c}{Claude} \\
    \cmidrule(lr){2-6}\cmidrule(lr){7-11}
    Rule & \textit{n} & Impr. & Gain & Cov. & Regr. & \textit{n} & Impr. & Gain & Cov. & Regr. \\
    \midrule
    A anchor only & \EThreeAN & \EThreeAImp & \EThreeAImpMed & \EHidGptACov & \EHidGptAReg & \ECThreeAN & \ECThreeAImp & \ECThreeAImpMed & \EHidCACov & \EHidCAReg \\
    R random & \EThreeRN & \EThreeRImp & \EThreeRImpMed & \EHidGptRCov & \EHidGptRReg & \ECThreeRN & \ECThreeRImp & \ECThreeRImpMed & \EHidCRCov & \EHidCRReg \\
    B boundary & \EThreeBN & \EThreeBImp & \EThreeBImpMed & \EHidGptBCov & \EHidGptBReg & \ECThreeBN & \ECThreeBImp & \ECThreeBImpMed & \EHidCBCov & \EHidCBReg \\
    C risk & \EThreeCN & \EThreeCImp & \EThreeCImpMed & \EHidGptCCov & \EHidGptCReg & \ECThreeCN & \ECThreeCImp & \ECThreeCImpMed & \EHidCCCov & \EHidCCReg \\
    C+F3 & \EThreeCFThreeN & \EThreeCFThreeImp & \EThreeCFThreeImpMed & -- & -- & -- & -- & -- & -- & -- \\
    P policy & \EThreePGptN & \EThreePGptImp & \EThreePGptImpMed & \EHidGptPGptCov & \EHidGptPGptReg & \ECThreePGptN & \ECThreePGptImp & \ECThreePGptImpMed & \EHidCPGptCov & \EHidCPGptReg \\
    P random & \EThreePRandN & \EThreePRandImp & \EThreePRandImpMed & -- & -- & \ECThreePRandN & \ECThreePRandImp & \ECThreePRandImpMed & -- & -- \\
    \bottomrule
  \end{tabular}
  \vspace{-6pt}
\end{table}

\section{From Verdicts to a Local Model: Family Labels Train the Better Policy}\vspace{-3pt}
\label{sec:local}
\looseness=-1
The training data come from seven textbook circuits at two to four sizings each, all laid out by one generic textbook generator, so the evaluation generators are never seen in training.
Edits start from the original source, a variant of it, or a copy with one injected fault; candidate patches come from deterministic rules and from Claude Opus~5.5 as a teacher, and each enters the data only after physical verification on its anchor and eight further configurations.
We fine-tune Qwen2.5-Coder-32B-Instruct~\cite{qwen2024coder} with LoRA~\cite{hu2022lora} under two labels: point supervision ranks the candidate patches of one starting source by the anchor alone, and family supervision ranks them by coverage, regression and quality across the family.
Each label is trained by supervised fine-tuning (SFT) and by direct preference optimization (DPO)~\cite{rafailov2023dpo} with matched data, budget and three seeds; the four models and the untuned base model then replace the frontier policy under rule C on one task per evaluation generator, and the hidden configurations are evaluated only after the training data are frozen.

\looseness=-1
Family labels train the better model.
On the hidden configurations, the model trained on family labels reaches $\Stwo$ on \EFourFamilyGain{} percentage points more than the same model trained on point labels, under SFT and DPO alike (95\% bootstrap interval \EFourFamilyCI{}), and both leave the untuned base model far behind (Fig.~\ref{fig:editing}(d)).
The family-trained models also regress least and improve the most tasks.
In \EFourFamilyFlips\% of the preference pairs the two labels disagree on which patch is better, and these pairs are where the two models part ways.
Supervision was the scarce resource: free-form plans from a frontier teacher produced no patch that applied to the source in forty attempts, while requests for the exact source text to replace did, and the final data hold \EFourContexts{} SFT examples, \EFourInjected{} of them repairs of injected faults, and \EFourPairs{} preference pairs.

\looseness=-1
Physical verification costs the same whichever model proposes the edits, about \EFiveGptPhysicalCpuH{} CPU hours for the GPT-6.1 searches and \EFiveClaudePhysicalCpuH{} for the Claude searches; parameter-baseline searches cost the most, since every setting they try is a full evaluation (Fig.~\ref{fig:editing}(e)).
Inference is what the local model saves: the Claude searches cost \EFiveClaudeUSD{} USD at list price, a local search costs \EFiveLocalPerSearchUSD{} USD of GPU time against \EFiveFrontierPerSearchUSD{} USD of frontier inference at the same hidden coverage, and the \EFiveTrainGPUh{} GPU-hours of training pay for themselves after \EFiveBreakEven{} searches.

\section{Discussion and Conclusion}\vspace{-3pt}
\label{sec:discussion}
\looseness=-1
Although a model can draw a layout from scratch, it is more valuable, more efficient and less costly when it works inside the generator: the code behind the layout is where its work is reused, checked and learned from.
A drawn layout is paid for in tokens and time at every new set of device sizes; a generator edit is paid for once and, if it holds across the family, serves every layout the generator produces.
The framework also makes that edit cheap to write and to trust: verified cells keep edits small, and because the generator is parameterized, one edit can be tested on other configurations before it is accepted.
The results of these family checks also serve as training labels, which let a smaller local model take over the editing at a fraction of the inference cost.
AI agents for analog layout become more cost-efficient on a well-defined generator framework, where each verified change becomes design knowledge that later configurations inherit.

\newpage% references start on their own page
\IEEEtriggeratref{13}% balance the two reference columns on the last page
\bibliographystyle{IEEEtran}
\bibliography{references}
\end{document}

%% file: numbers.tex
\newcommand{\EOneGZeroN}{119}

\newcommand{\EOneGOneN}{118}

\newcommand{\EOneGTwoReuseN}{52}
\newcommand{\EOneGTwoReuseS}{44}

\newcommand{\EOneSuccessHiddenFull}{45}

\newcommand{\EOneGZeroFirstDiffLines}{32}

\newcommand{\EOneGOneFirstDiffLines}{41}

\newcommand{\EOneGTwoOutTok}{242}

\newcommand{\EOneGTwoMedianDecisions}{5}

\newcommand{\EOneGTwoTokPerVerifiedK}{0.3}
\newcommand{\EOneGTwoCpuPerVerifiedMin}{4}

\newcommand{\EOneLowTokPerVerifiedK}{16.1}
\newcommand{\EOneLowCpuPerVerifiedMin}{3}

\newcommand{\EOneLowToGTwoTokRatio}{56}
\newcommand{\EOneSuccessTotal}{46}

\newcommand{\ETwoSZeroSpearman}{-0.97}

\newcommand{\ETwoMillerSZeroUgbSpread}{0.04}

\newcommand{\ETwoOnMillerSZeroAreaSTwo}{4}

\newcommand{\EThreeAN}{120}
\newcommand{\EThreeAImp}{37}

\newcommand{\EThreeAImpMed}{39}

\newcommand{\EThreeRN}{120}
\newcommand{\EThreeRImp}{25}

\newcommand{\EThreeRImpMed}{48}

\newcommand{\EThreeBN}{120}
\newcommand{\EThreeBImp}{23}

\newcommand{\EThreeBImpMed}{48}

\newcommand{\EThreeCN}{120}
\newcommand{\EThreeCImp}{15}

\newcommand{\EThreeCImpMed}{48}

\newcommand{\EThreeCFThreeN}{10}
\newcommand{\EThreeCFThreeImp}{3}

\newcommand{\EThreeCFThreeImpMed}{119}

\newcommand{\EThreePGptN}{40}
\newcommand{\EThreePGptImp}{31}

\newcommand{\EThreePGptImpMed}{39}

\newcommand{\EThreePRandN}{40}
\newcommand{\EThreePRandImp}{25}

\newcommand{\EThreePRandImpMed}{13}

\newcommand{\EThreeSiblingFailures}{43}

\newcommand{\ECThreeSiblingFailures}{12}

\newcommand{\ECThreeAN}{72}
\newcommand{\ECThreeAImp}{36}

\newcommand{\ECThreeAImpMed}{62}
\newcommand{\ECThreeRN}{72}
\newcommand{\ECThreeRImp}{27}

\newcommand{\ECThreeRImpMed}{48}
\newcommand{\ECThreeBN}{72}
\newcommand{\ECThreeBImp}{32}

\newcommand{\ECThreeBImpMed}{48}
\newcommand{\ECThreeCN}{72}
\newcommand{\ECThreeCImp}{29}

\newcommand{\ECThreeCImpMed}{45}

\newcommand{\ECThreePGptN}{12}
\newcommand{\ECThreePGptImp}{12}

\newcommand{\ECThreePGptImpMed}{38}
\newcommand{\ECThreePRandN}{12}
\newcommand{\ECThreePRandImp}{11}

\newcommand{\ECThreePRandImpMed}{12}

\newcommand{\ECOneSuccessTotal}{84}
\newcommand{\ECOneHiddenFull}{82}

\newcommand{\ECOneLowToGTwoTokRatio}{2}

\newcommand{\ECThreeCompImpMax}{80}

\newcommand{\ECThreeOtaImpMax}{352}

\newcommand{\EFiveGptPhysicalCpuH}{148}

\newcommand{\EFiveClaudeUSD}{382}
\newcommand{\EFiveClaudePhysicalCpuH}{106}